\documentstyle[aps]{revtex}
\font\eightrm=cmr8

\def\bt0zk{{\scriptstyle({\eightrm t}_{\eightrm 0}, {\eightrm{\bf z}})}}
\def\mmbox#1#2{\vcenter{\hrule \hbox{\vrule height#2in
                \kern#1in \vrule} \hrule}}
\def\dAlemb{\mmbox{.09}{.09}}
\input{epsf}
\begin{document}
\draft
\begin{flushright} TOKAI-HEP/TH-0007 \end{flushright}
\vspace{0.5cm}
\centerline{\bf Relativistically covariant formulation}
\centerline{\bf of the canonical theory of classical fields II}
\vspace{1.0cm}
\centerline{Hiroshi Ozaki\footnote{Email address: ozaki@keyaki.cc.u-tokai.ac.jp}}
\vspace{0.5cm}
\centerline{\small {\it Department of Physics, Tokai University, 1117 Kitakaname, Hiratsuka 259-1292, Japan}}

\begin{abstract}
A covariant description of the canonical theory for interacting classical fields is developed on a
space-like hypersurface. An identity invariant under the canonical transformations is
obtained. The identity follows a canonical equation in which  
the interaction Hamiltonian density genarates a deformation of the space-like hypersurface.
The equation just corresponds to the Yang-Feldman equation in the Heisenberg
pictures in quantum field theory.

\end{abstract}
\pacs{ }


\section{Introduction}

In a previous paper~\cite{oz1} (to be refered to in what follows as I) we built up a canonical formalism 
of classical fields on a space-like hypersurface.
A functional derivative on a space-like hypersurface, Hamilton's equations on the hypersurface,
and four-dimensional Poisson bracket were introduced to make the canonical formalism
free from restriction of the equal-time. 
The canonical formalism of this kind was attempted by Koba~\cite{Koba} about 50 years ago.
At that time he introduced dynamical variable $\xi(x;\sigma)$, depending on a space-time point $x$ and a space-like
hypersurface $\sigma$, and claimed that the responce law of $\xi(x;\sigma)$ 
under the deformation of $\sigma$ is given by the interaction Hamiltonian density $H(x';n(x'))$:
\begin{equation}
{{\delta \xi (x;\sigma )} \over {\delta \sigma (x')}}
=\left[ {\xi (x;\sigma ),H(x';n(x'))} \right]_{{\rm P.B.}}+{{\partial \xi (x;\sigma )} \over {\partial \sigma (x')}},
\label{eq:IR1}
\end{equation}
where the subscript P.B. means a covariant Poisson bracket and $n(x')$ is the normal vector on a point $x'$.
Further, he assumed the covariant Poisson bracket satisfies all algebraic relations of the ordinary three-dimensional Poisson
bracket and set $\left[ {\xi (x;\sigma ),\xi (y;\sigma )} \right]_{{\rm P.B.}}$ to be proportional
to an invariant delta function. 
The integravility condition of (\ref{eq:IR1}):
\begin{equation}
{\delta  \over {\delta \sigma (x'')}}{{\delta \xi (x;\sigma )} \over {\delta \sigma (x')}}
={\delta  \over {\delta \sigma (x')}}{{\delta \xi (x;\sigma )} \over {\delta \sigma (x'')}},
\end{equation}
which leads to the Poisson bracket relation
\begin{equation}
\left[ {H(x';n(x')),H(x'';n(x''))} \right]_{{\rm P.B.}}-
{{\partial H(x'';n(x''))} \over {\partial \sigma (x')}}+
{{\partial H(x';n(x'))} \over {\partial \sigma (x'')}}={\rm const.}\ . 
\label{eq:IR2}
\end{equation}
This relation was used to define $H(x';n(x'))$; this is a generating function to deform $\sigma$.
He applied this formalism to the problem of finding the interaction Hamiltonian density
for the interacting scalar-meson photon field. However, there remained ambuguities on the definition
of $\xi (x;\sigma )$ and the covariant Poisson bracket. It was difficult to extend his prescription to
the general theory because of these ambuguities.
In this paper we will remove those ambuguities with tools introduced in I
and develop the general theory. It turns out that $\xi (x;\sigma )$ is constructed by the field equation
with an invariant delta function; the covariant Poisson bracket 
is realized by the four-dimensional Poisson bracket defined in I.
Further, it will be turned out that Koba's attempt is connected to geometrical approach
and we can pass from our canonical formalism to
the Yang-Feldman theory without spoiling the explicit Lorentz covariance.
Throughout this paper, notations
$\Lambda (\partial ),\ d(\partial ),\ 
\Gamma _\mu (\partial ,-\mathord{\buildrel{\lower3pt\hbox{$\scriptscriptstyle\leftarrow$}}\over \partial } )$
, and $\eta$ are according to Takahashi~\cite {Takahashi} (1969); $\partial_n$, $\partial_{t\mu}$, $P_{\mu\nu}$,
and $\bigl[F[\sigma], G[\sigma] \bigr]_c$ are according to I.

\section{The bubble differentiation}

\begin{figure}
\epsfxsize=4cm
\centerline{\epsfbox{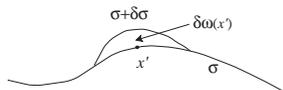}}
\caption{A deformation of the surface $\sigma$ at the point $x^{\prime}$.
The upper surface $\sigma +\delta\sigma$ and the lower surfaces $\sigma$ overlap everywhere
except a small region surounding the point $x^{\prime}$. The volume element $\delta\omega (x^{\prime})$ enclosed
between the surfaces is taken to approach zero around the point $x^{\prime}$.}
\label{fig1}
\end{figure}

The bubble differentiation with respect to super-many-time was first
introduced by Tomonaga~\cite{Tom}. 
Let us put a space-like hypersurface $\sigma$ by
a function of $x,y,z$, say $g(x,y,z)$, and a real parameter $c$:
\begin{equation}
\sigma =g(x,y,z)+c. \label{eq:BD1}
\end{equation}
Put a functional $F[\sigma ]$ on $\sigma$. 
If we deform $g(x,y,z)$ with $c$ fixed, the surface $\sigma$ changes into another surface. 
$F[\sigma]$ will also change into another functional under the deformation of $\sigma$.
Then a differentiation with respect to $\sigma$ is defined by~\cite {Tom,KTT,SW1} 
\begin{equation}
{{\delta F[\sigma ]} \over {\delta \sigma (x')}}\equiv 
\mathop {\lim }\limits_{\delta \omega (x')\to 0}{{F[\sigma ']-F[\sigma ]} \over {\delta \omega (x')}}, \label{eq:BD2}
\end{equation}
where $\sigma'$ is a surface overlapping $\sigma$ everywhere except a small region surrounding
the point $x'$, and $\delta\omega (x')$ is the volume element of the small world region enclosed by
$\sigma$ and $\sigma'$, as shown in Figure \ref{fig1}.
This is called the bubble differentiation.
Note that the difference between the bubble differentiation and the usual
time derivative. The latter is a derivative with respect to $c$ leaving $g(x,y,z)$ constant.

A special class of functional is
\begin{equation}
F[\sigma ]=\int_\sigma  {d\Sigma _\mu (x'')f^\mu [x'']}, \label{eq:BD3}
\end{equation}
that is a surface integral of a point function. The bubble differentiation of (\ref{eq:BD3})
gives the Lorentz scalar density:
\begin{equation}
{{\delta F[\sigma ]} \over {\delta \sigma (x')}}=\partial '_\mu f^{\mu} [x'], \label{eq:BD4}
\end{equation}
according to Gauss' Theorem
\begin{equation}
\int_\Omega  {d^4x''\partial'' _\mu f^\mu [x'']}=\int_{\partial \Omega } {d\Sigma _\mu (x'')f^\mu [x'']}. \label{eq:BD5}
\end{equation}
In the next section we will consider the role of the bubble differentiation (\ref{eq:BD4}) in the
canonical formalism.

\section{Canonical Identity}

Let us consider a multi-component classical field $\phi_{\alpha}(x'')$ 
($\alpha = 1, 2, \cdots, N$) at the point $x''$
on a space-like hypersurface $\sigma$  interacting with the source
$J[\phi_{\alpha}(x'')]$; this is a functional of the classical field and its derivatives.
We now assume the Lagrangian density ${\cal L}[x'']$ for the interacting field can be decomposed
into a kinetic term ${\cal L}_0 [x'']$ and an interaction term ${\cal L}_{\rm int} [x'']$.
Then the Euler-Lagrange equation reads
$$
\Lambda_{\alpha\beta} (\partial'')\phi _{\beta}(x'')=J[\phi _{\alpha}(x'')]
$$
or, in matrix notation,
\begin{equation}
\Lambda (\partial'')\phi (x'')=J[\phi (x'')]. \label{eq:CI1}
\end{equation}
Hereafter we adopt matrix notation and drop the index of $\phi (x'')$.
We assume the operator $\Lambda (\partial'')$ fulfils the following conditions~\cite{Takahashi}:
\begin{description}
	\item[{[i]}] There exists a non-singular matrix $\eta$ such that
	\begin{equation}
	\left[ {\eta \Lambda (\partial'' )} \right]^\dagger =\eta \Lambda (-\partial'' ). \label{eq:CI2}
	\end{equation}
	\item[{[ii]}] $\Lambda (\partial'')$ is a polynomial of the derivative operator $\partial''_{\mu}$:
	\begin{equation}
	\Lambda (\partial'' )=\Lambda _0+\Lambda _{\mu }\partial ^{''\mu} +
	\Lambda _{\mu \nu }\partial ^{''\mu} \partial ^{''\nu} , \label{eq:CI3}
	\end{equation}
	where the coefficients $\Lambda_{0}, \Lambda_{\mu}, {\rm and} \ \Lambda_{\mu\nu}$ do not depend on $x''$.
\end{description}
The former assumption is equivalent to express ${\cal L}_0 [x'']$ as
$$
{\cal L}_0 [x'']=\bar \phi (x'')\Lambda (\partial'')\phi (x'')
$$
from which the field equation (\ref{eq:CI1}) and the adjoint field equation
\begin{equation}
\bar \phi (x'')\Lambda (-\mathord{\buildrel{\lower3pt\hbox{$\scriptscriptstyle\leftarrow$}}\over {\partial''} } )
=(\eta J[\phi (x'')])^\dagger \label{eq:CI4}
\end{equation}
with
$$
\bar \phi (x'')=\phi ^\dagger (x'')\eta 
$$
are derivable.

At any point $x$ in space-time,
the operator $\Lambda(\partial)$ gives the Klein-Gordon operator ($\dAlemb + m^2$) with the
Klein-Gordon divisor $d(\partial)$:
\begin{equation} 
\Lambda (\partial )d(\partial )=d(\partial )\Lambda (\partial )=\varepsilon(\dAlemb + m^2), \label{eq:CI5}
\end{equation}
where
$$
\varepsilon=\left\{\begin{array}{ll}
\ 1 & \mbox{excepting gauge fields} \\
-1 & \mbox{gauge fields.}
\end{array}\right.
$$
The gauge fields need the factor $\varepsilon$ because of including a negative sign in ${\cal L}_0 [x'']$.
The Klein-Gordon operator accompanies with an invariant delta function $\Delta (x)$:
\begin{equation}
(\dAlemb + m^{2}) \Delta (x) =0. \label{eq:CI6}
\end{equation}

On the other hand, according to I, the interacting field on $\sigma$ satisfies Hamilton's equations
\begin{eqnarray}
(-)^{\left| \phi \right|+1}\partial ''_n\phi (x'')-\partial ''_{t\mu }
{{\partial ^R{\cal H}[x'']} \over {\partial [\partial'' _{t\mu }\Pi (x'')]}}
+{{\partial ^R{\cal H}[x'']} \over {\partial \Pi (x'')}}&=&0, \label{eq:CI7a}\\
  \partial'' _n\Pi (x'')-\partial'' _{t\mu }{{\partial ^{R}{\cal H}[x'']} \over {\partial [\partial'' _{t\mu }\phi (x'')]}}
  +{{\partial ^{R}{\cal H}[x'']} \over {\partial \phi (x'')}}&=& 0 \label{eq:CI7b}.
\end{eqnarray}
The $\vert \phi \vert$ is the number factor associated with the classical field:
take $0$ for Grassmann even function, and $1$ for Grassmann odd function.
The superscript ^^ ^^ $\ R$ " means that we have adopted the right-differentiation convention.

We assume the Hamiltonian density ${\cal H}[x'']$ for the interacting field can also be decomposed
into a kinetic term ${\cal H}_0 [x'']$ and an interaction term ${\cal H}_{\rm int} [x'']$.
Our assumption on ${\cal L}[x'']$ and ${\cal H}[x'']$ 
allows us to extract the kinetic term from the Euler-Lagrange and Hamilton's equations, respectively.
Using the adjoint field equation, we have
\begin{eqnarray}
\Lambda (\partial'' )\phi (x'')&=&(-)^{\left| \phi \right| }\eta ^{-1}\left[ {\partial'' _n\Pi _{0}(x'')-\partial'' _{t\mu }
{{\partial ^{R}{\cal H}_0[x'']} \over {\partial [\partial'' _{t\mu }\phi (x'')]}}
+{{\partial ^{R}{\cal H}_0[x'']} \over {\partial \phi (x'')}}} \right]^\dagger \nonumber \\
  &=&\Lambda _0\phi (x'')+\Lambda _\mu \partial ^{''\mu} \phi (x'')+\Lambda _{\mu \nu }\partial^{''\mu} \partial ^{''\nu} \phi (x''),
  \label{eq:CI8}
\end{eqnarray}
where
$$
\Pi _{0}(x'')=n_\mu (x''){{\partial ^R{\cal L}_0[x'']} \over {\partial [\partial ''_\mu \phi (x'')]}}.
$$
The coefficient $\Lambda_0$ does not contain the derivative operator $\partial''_{\mu}$, so that
\begin{eqnarray*}
 (-)^{\left| \phi \right| }\eta ^{-1}\left( {{{\partial ^{R}{\cal H}_0[x'']} 
\over {\partial \phi (x'')}}} \right)^\dagger&=&\Lambda _0\phi (x'') \nonumber \\
 (-)^{\left| \phi \right| }\eta ^{-1}\left( {\partial'' _n\Pi _0 (x'')-\partial'' _{t\mu }{{\partial^{R}{\cal H}_0[x'']} 
  \over {\partial [\partial'' _{t\mu }\phi (x'')]}}} \right)^{\dagger}&=&\Lambda _\mu \partial ^{''\mu} \phi (x'')
  +\Lambda _{\mu \nu }\partial^{''\mu} \partial^{''\nu} \phi (x''). 
\end{eqnarray*}

Define
\begin{eqnarray*}
S^\nu (x'')&=&n^\nu (x'')\Pi _{\rm int}(x'')
-P^{\nu \mu }{{\partial ^R{\cal H}_{\rm int}[x'']} \over {\partial [{\partial'' _t}^{\mu} \phi (x'')]}}, \\
  \Pi _{\rm int}(x'')&=&n_\mu (x''){{\partial ^{R}{\cal L}_{\rm int}[x'']} \over {\partial [\partial ''_\mu \phi (x'')]}}, 
\end{eqnarray*}
and
\begin{equation}
\Gamma ^\nu (\partial '',\mathord{\buildrel{\lower3pt\hbox{$\scriptscriptstyle\leftarrow$}}\over {-\partial''} })
=\Lambda ^\nu +\Lambda ^{\nu \lambda }(\partial ''_\lambda 
-\mathord{\buildrel{\lower3pt\hbox{$\scriptscriptstyle\leftarrow$}}\over {\partial''} }_\lambda ) \nonumber.
\end{equation}
Then a fundamental identity for the interacting field can be derived 
quite simply by use of (\ref{eq:CI6}) and (\ref{eq:CI7b}):
\begin{eqnarray}
0&=&d(\partial )\Lambda (\partial )\Delta (x-x'')\cdot \phi (x'') \nonumber \\ 
 & &{ } - (-)^{\left| \phi \right| }d(\partial )\Delta (x-x'')\!\cdot\! \eta ^{-1}\left( {\partial'' _n\Pi (x'')-\partial'' _{t\mu }
  {{\partial ^R {\cal H}[x'']} \over {\partial [\partial'' _{t\mu }\phi (x'')]}}
  +{{\partial ^R {\cal H}[x'']} \over {\partial \phi (x'')}}} \right)^\dagger \nonumber \\
  &=&-\partial ''_\nu \left[ {d(\partial )\Delta (x-x'') \!\cdot\! 
  \left( {\Gamma ^\nu (\partial '',\mathord{\buildrel{\lower3pt\hbox{$\scriptscriptstyle\leftarrow$}}\over {-\partial''} })
  \phi (x'')+ (-)^{\left| \phi \right| }\eta ^{-1}{S^\nu}^\dagger (x'')} \right)} \right] \nonumber \\
  & &{ }+ (-)^{\left| \phi \right| }\partial ''_\nu d(\partial )\Delta (x-x'')\!\cdot\! \eta ^{-1} {S^\nu}^\dagger (x'')
  - (-)^{\left| \phi \right| }d(\partial )\Delta (x-x'')\!\cdot\! 
  \eta ^{-1}\left( {{{\partial ^{R}{\cal H}_{\rm int}[x'']} \over {\partial \phi (x'')}}} \right)^\dagger,  \nonumber \\
  & &\label{eq:CI9}
\end{eqnarray}
The identity remains invariant under the canonical transformations since the Klein-Gordon equation
(\ref{eq:CI6}) does not depend on the choice of canonical variables
and the canonical equation (\ref{eq:CI7b}) holds for any canonical variables. 
Thus we will call (\ref{eq:CI9}) the canonical identity.
Note that the point $x$ is not necessarily on the surface $\sigma$, but the point $x''$ is on $\sigma$.

Let us define a surface integral:
\begin{equation}
\phi (x;\sigma )\equiv - \int_\sigma  
{d\Sigma _\nu (x'')d(\partial )\Delta (x-x'')\cdot }
\left[ {\Gamma ^\nu (\partial '',\mathord{\buildrel{\lower3pt\hbox{$\scriptscriptstyle\leftarrow$}}\over {-\partial''} } )
\phi (x'')+ (-)^{\left| \phi \right| }\eta ^{-1}{S^\nu}^{\dagger} (x'')} \right]. \label{eq:CI10}
\end{equation}
With the help of (\ref{eq:BD4}), we have immediately
\begin{eqnarray}
{{\delta \phi (x;\sigma )} \over {\delta \sigma (x')}}&=&
- (-)^{\left| \phi \right| }\partial '_\nu d (\partial )\Delta (x-x')\!\cdot\! \eta ^{-1}{S^{\nu }}^\dagger (x') \nonumber\\
& &\qquad+ (-)^{\left| \phi \right| }d (\partial )\Delta (x-x')\!\cdot\! \eta ^{-1}
\left( {{{\partial ^R{\cal H}_{\rm int}[x']} \over {\partial \phi (x')}}} \right)^\dagger . \label{eq:CI10b}
\end{eqnarray}
This is an alternative of the canonical identity (\ref{eq:CI9}), so it is a canonical invariant quantity.

The general properties of $\phi (x;\sigma)$ are
\begin{enumerate}

\item   	$\phi (x;\sigma)$ satisfies the free-field equation:
		\begin{equation}
		\Lambda (\partial )\phi (x;\sigma )=0. \label{eq:CI11}
		\end{equation}
\item		The four-dimensional Poisson bracket of $\phi (x;\sigma )$ and $\bar\phi  (y;\sigma )$ is
		 \begin{equation}
		 \left[ \phi (x;\sigma ), \bar\phi (y;\sigma ) \right]_c=  d (\partial_x )\Delta (x-y). \label{eq:CI12}
		 \end{equation}
\end{enumerate}
		To verify (\ref{eq:CI11}) operate $\Lambda(\partial)$ from the left on $\phi (x;\sigma )$. Then
		\begin{eqnarray*}
		\Lambda (\partial )\phi (x;\sigma ) &=&\!\!
		\int_\sigma  {d\Sigma _\nu (x'')\Lambda (\partial )d(\partial )\Delta (x-x'')\cdot }
		\left[ {\Gamma ^\nu 
		(\partial '',\mathord{\buildrel{\lower3pt\hbox{$\scriptscriptstyle\leftarrow$}}\over {-\partial''} } )
		\phi (x'')+(-)^{\left| \phi \right| }\eta ^{-1}{S^\nu}^{\dagger} (x'')} \right] \\
 		 &=&\!\!\int_\sigma  {d\Sigma _\nu (x'')
  		 {(\dAlemb+m^2)\Delta (x-x'')\!\cdot\! 
  		\left[ {\Gamma ^\nu 
  		(\partial '',\mathord{\buildrel{\lower3pt\hbox{$\scriptscriptstyle\leftarrow$}}\over {-\partial''} } )
  		\phi (x'')+(-)^{\left| \phi \right| }\eta ^{-1}{S^\nu}^{\dagger} (x'')} \right]} } \\
  		&=& 0.
  		\end{eqnarray*}

		 To verify (\ref{eq:CI12}) we need to know whether the classical field is a c-number function or
		 a anticommuting c-number function.\par
\vspace{5mm}
\noindent
{\bf[ case (a)]} The c-number field \par\noindent
		Since the c-number field (classical Bose field: $\left| \phi \right| =0$) satisfies a second-order field equation,
		the field equation of the c-number field do not contain
		the first derivative of $\phi (x'')$ such as $\Lambda _\mu {\partial''}^\mu \phi (x'')$,
		and the non-sigular matrix $\eta$ reads the unit matrix or unity. 
		Thus we have
		\begin{eqnarray*}
		\Lambda (\partial '')\phi (x'')&=&
		\Lambda _0\phi (x'')+\Lambda _{\mu \nu }\partial ''^\mu \partial ''^\nu \phi (x''), \\
		\eta ^{-1}&=&1.
		\end{eqnarray*}
		Hence
		\begin{eqnarray*}
		\phi (x;\sigma )&=&
		\int_\sigma  d\Sigma 
		\Bigl[ n_\nu (x'')\Lambda ^{\nu \lambda }
		\partial ''_\lambda d (\partial )\Delta (x-x'')\cdot \phi (x'')  \\
		& &{ }\qquad\qquad -  d (\partial )\Delta (x-x'')\cdot \Pi^\dagger (x'') \Bigr].
		\end{eqnarray*}
		and
		\begin{eqnarray*}
		\bar \phi (x;\sigma ) &=& \phi ^\dagger (x;\sigma )  \\
		 &=&\int_\sigma  d\Sigma \Bigl[ n_\nu (x'')
		 {\Lambda ^{\nu \lambda }} \partial ''_\lambda d (-\partial )\Delta (x-x'')\cdot \phi ^\dagger (x'')
		 \nonumber \\
  		 & &\qquad\qquad -  d (-\partial )\Delta (x-x'')\cdot \Pi (x'') \Bigr],
		\end{eqnarray*}
		the last step being verified with the property of the Klein-Gordon divisor~\cite{Takahashi}:
		$\left[ {d(\partial )\eta ^{-1}} \right]^\dagger =d(-\partial )\eta ^{-1}$.
		For any two points $x$ and $y$ in space-time we have
		\begin{eqnarray*}
		\left[ {\phi (x;\sigma ), \bar \phi (y;\sigma )} \right]_c 
		&=&
		\int_\sigma  d\Sigma 
		\left( {{\tilde \delta \phi (x;\sigma )} \over {\tilde \delta \phi (x')}}
		{{\tilde \delta  \bar \phi (y;\sigma )} \over {\tilde \delta  \Pi (x')}}-
		{{\tilde \delta \phi (x;\sigma )} \over {\tilde \delta  \Pi (x')}}
		{{\tilde \delta  \bar \phi (y;\sigma )} \over {\tilde \delta \phi (x')}} \right)  \\
		  &=&\!\!\int_\sigma  d\Sigma 
		  \Bigl[ -n_\nu (x')\Lambda ^{\nu \lambda }\partial '_\lambda d (\partial_x )\Delta (x-x')
		  \!\cdot\! d (-\partial_y )\Delta (y-x')  \\
		  & &\quad+d (\partial_x )\Delta (x-x')\cdot
		   n_\nu (x')\Lambda ^{\nu \lambda }\partial '_\lambda d (-\partial_y )\Delta (y-x') \Bigr]\nonumber \\
		 &=&
		  d (-\partial_y )\Delta (x-y) \\
		 &=&
		   d (\partial_x )\Delta (x-y).
		\end{eqnarray*}
\noindent
{\bf[case (b)]} The anticommuting c-number field \par\noindent
	        Each component of the anticommuting c-number field (classical Fermi field :$\left| \phi \right| =1$)
		is taken to be an element of an infinite-dimensional Grassmann algebra.
		Since the anticommuting c-number field satisfies a first-order field equation, 
		the field equation of classical Fermi field does not contain the second order derivative
		term such as $\Lambda _{\mu \nu }\partial ''^\mu \partial ''^\nu \phi (x'')$.
		Hence
		\begin{eqnarray*}		
		\phi (x;\sigma )&=&\int_\sigma  
		{d\Sigma \ d (\partial )\Delta (x-x'')\cdot \eta ^{-1}\Pi ^\dagger (x'')}\\
		  &=&
		  -\int_\sigma  
		{d\Sigma \ d (\partial )\Delta (x-x'')\cdot 
		\left[ {n_\nu (x'')\Lambda ^\nu \phi (x'')-\eta ^{-1}\Pi _{\rm int} ^\dagger (x'')} \right]}
		\end{eqnarray*}
		and
		\begin{eqnarray*}
		\bar \phi (x;\sigma )&=&
		 \int_\sigma    
		{d\Sigma \left[ {\ d (\partial )\Delta (x-x'')\cdot \eta ^{-1}\Pi ^\dagger (x'')} \right]}^\dagger 
		\eta \\
		&=&
		\int_\sigma  {d\Sigma \ \Pi (x'')d (-\partial)\Delta (x-x'')}.
		\end{eqnarray*}
		Note that we take a simplified treatment in which $\bar\phi (x'')$ is considered
		as a momentum $\Pi (x'')$ conjugate to $\phi (x'')$.
		Then we have
		\begin{eqnarray*}
		\left[ {\phi (x;\sigma ),\bar \phi (y;\sigma )} \right]_c &=&
		\int_\sigma  
		{d\Sigma \left( {
		{{\tilde \delta      \phi (x;\sigma )} \over {\tilde \delta \phi (x')}}
		{{\tilde \delta  \bar\phi (y;\sigma )} \over {\tilde \delta \Pi (x')}}+
		{{\tilde \delta      \phi (x;\sigma )} \over {\tilde \delta \Pi (x')}}
		{{\tilde \delta  \bar\phi (y;\sigma )} \over {\tilde \delta \phi (x')}}
		 }\right)} \\
		&=&\int_\sigma  \ {d\Sigma \left[ {(-d (\partial_x ))\Delta (x-x') n_\nu (x')\Lambda ^\nu \cdot 
		d (-\partial_y )\Delta (y-x')} \right]} \\
		&=&d (\partial_x )\Delta (x-y).
		\end{eqnarray*}
%
%
Thus the four-dimensional Poisson bracket of $\phi(x;\sigma)$ and $\bar\phi(y;\sigma)$ is
proportional to the invariant delta function $\Delta (x-y)$. This shows that
the covariant Poisson bracket proposed by Koba~\cite{Koba} was realized by the four-dimensional Poisson bracket.
		
If there is no interaction, we have ${\cal L}_{\rm int}[x'']={\cal H}_{\rm int}[x'']= 0 $. 
Then $\phi (x; \sigma)$ become
\begin{equation}
\phi (x;\sigma )=
-\int_\sigma  {d\Sigma _\nu (x'')\left[ {d (\partial )\Delta (x-x'')\cdot 
\Gamma ^\nu (\partial '',\mathord{\buildrel{\lower3pt\hbox{$\scriptscriptstyle\leftarrow$}}\over {-\partial''} } )
\phi (x'')} \right]}. \label{eq:CI13}
\end{equation}
On the other hand,  
the solution of the Cauchy problem for the free-field equation:
\begin{equation} 
\Lambda (\partial) \phi (x) = 0 \nonumber
\end{equation}
are given by~\cite{Takahashi}
\begin{equation}
\phi (x)=
-\int_\sigma  {d\Sigma _\nu (x'')\left[ {d (\partial )\Delta (x-x'')\cdot 
\Gamma ^\nu (\partial '',\mathord{\buildrel{\lower3pt\hbox{$\scriptscriptstyle\leftarrow$}}\over {-\partial''} } )
\phi (x'')} \right]}. \label{eq:CI14a}
\end{equation} 
Thus $\phi (x, \sigma)$ reduces to $\phi (x)$ for the free field. 
It follows that
\begin{equation}
\left[ {\phi (x),\bar \phi (y)} \right]_c =  d (\partial_x )\Delta (x-y), \label{eq:CI14b}
\end{equation}
which corresponds to the four-dimensional commutation relation between components of 
the free field in quantum field theory.

\section{Canonical Equation}

Let us go back to (\ref{eq:CI10}) for the interacting field.
It can be also related to the four-dimensional Poisson bracket of $\phi (x; \sigma)$  and
${\cal H}_{\rm int} [x']$.
For the c-number field, with the help of $(A.1)$, we have
\begin{eqnarray}
{{\delta \phi (x;\sigma )} \over {\delta \sigma (x')}} &=&
\left[ {\phi (x;\sigma ),{\cal H}_{\rm int}[x']} \right]_c \nonumber\\
  & & { }- \partial '_nd (\partial )\Delta (x-x')\cdot \Pi_{\rm int}^\dagger (x')\nonumber \\
  & & { }-n_\nu (x')\Lambda ^{\nu \lambda }\partial '_\lambda 
  d (\partial )\Delta (x-x')\!\cdot\! 
  \left( {{{\partial ^R{\cal H}_{\rm int}[x']} \over {\partial \Pi(x')}}+
  \mathord{\buildrel{\lower3pt\hbox{$\scriptscriptstyle\leftarrow$}}\over {{\partial'_\mu}^{t}} }
  {{\partial ^R{\cal H}_{\rm int}[x']} \over {\partial [{\partial '_\mu}^t\Pi (x')]}}} \right)\nonumber \\
  &=&
  -\partial '_\nu d (\partial )\Delta (x-x')\cdot {S^{\nu }}^\dagger (x')
     + d (\partial )\Delta (x-x')\cdot
     \left({{{\partial ^R{\cal H}_{\rm int}[x']} \over {\partial \phi (x')}}} \right)^{\dagger}. \nonumber\\
  & & \label{eq:CI16}
\end{eqnarray}
To verify (\ref{eq:CI16}) we assume the hermiticity of ${\cal H}_{\rm int}[x']$ where 
$\phi (x')$ and $\bar\phi (x')$ appear symmetrically.
It follows that
$$
\left( {{{\partial ^R{\cal H}_{\rm {int}}[x']} \over {\partial \phi (x')}}} \right)^\dagger 
={{\partial ^R{\cal H}_{\rm {int}}[x']} \over {\partial \phi ^\dagger (x')}},
$$
and
$$
\left( {{{\partial ^R{\cal H}_{\rm {int}}[x']} \over  \partial [{\partial '_{t\mu}}\phi (x')]}} \right)^\dagger 
={{\partial ^R{\cal H}_{\rm {int}}[x']} \over  \partial [{\partial '_{t\mu}}\phi ^\dagger(x')]}.
$$
If we take a complex scalar field as an example,
these relations will be easily verified.
Thus (\ref{eq:CI16}) holds for not only the c-number hermitian field but also the c-number non-hermitian field.

For the anticommuting c-number field, with the help of $(A.1)$--$(A.3)$, we have
\begin{eqnarray}
{{\delta \phi (x;\sigma )} \over {\delta \sigma (x')}} &=&
-\left[ {\phi (x;\sigma ),{\cal H}_{\rm int}[x']} \right]_c \nonumber \\
  & &{ }-\partial '_\nu \!\left[ {d (\partial )\Delta (x-x')\!\cdot\! 
  \left( -{\eta ^{-1}{S^\nu }^\dagger (x')+P^{\nu \mu }
  \zeta\!\cdot \!
  {{\partial ^R{\cal H}_{\rm int}[x']} \over {\partial [{\partial '_t}^\mu \Pi (x')]}}} \right)} \right] \nonumber \\
  &=& \partial '_\nu d (\partial )\Delta (x-x')\!\cdot\! 
  \eta ^{-1}{S^{\nu }}^\dagger (x')-d (\partial )\Delta (x-x')\!\cdot\! 
  \eta ^{-1}\left( {{{\partial ^R{\cal H}_{\rm int}[x']} \over {\partial \phi (x')}}} \right)^\dagger , \nonumber \\
  & &\label{eq:CI17}
\end{eqnarray}
where 
$$
\zeta =-\left( {{{\partial ^R} \over {\partial \phi (x')}}
-\partial' _{t\mu }{{\partial ^R} \over {\partial [\partial' _{t\mu }\phi (x')]}}} \right)
\eta ^{-1}\Pi ^\dagger (x').
$$
The additive terms to $\left[ {\phi (x;\sigma ),{\cal H}_{\rm int}[x']} \right]_c$
in (\ref{eq:CI16}) and (\ref{eq:CI17}) are canonical invariant quantities since the canonical identity (\ref{eq:CI9})
and the four-dimensional Poisson bracket remain invariant under the canonical transformations.
Further, these additive terms nearly always vernish for ${\cal L}_{\rm int} [x']$ and ${\cal H}_{\rm int} [x']$
in practical problems for interacting fields:
\begin{eqnarray}
  & & { } \partial '_nd (\partial )\Delta (x-x')\cdot \Pi_{\rm int}^\dagger (x')\nonumber \\
  & & { }+n_\nu (x')\Lambda ^{\nu \lambda }\partial '_\lambda 
  d (\partial )\Delta (x-x')\!\cdot\! 
  \left( {{{\partial ^R{\cal H}_{\rm int}[x']} \over {\partial \Pi(x')}}+
  \mathord{\buildrel{\lower3pt\hbox{$\scriptscriptstyle\leftarrow$}}\over {{\partial'_{t\mu}}} }
  {{\partial ^R{\cal H}_{\rm int}[x']} \over {\partial [{\partial '_\mu}^t\Pi (x')]}}} \right) =0 \nonumber\\
  & &\label{eq:CI18}
\end{eqnarray}
for the c-number field, and
\begin{equation}
  \partial '_\nu \!\left[ {d (\partial )\Delta (x-x')\!\cdot\!\! 
  \left( -{\eta ^{-1}{S^\nu }^\dagger (x')+P^{\nu \mu }
  \zeta\!\cdot \!
  {{\partial ^R{\cal H}_{\rm int}[x']} \over {\partial [{\partial '_t}^\mu \Pi (x')]}}} \right)} \right]=0 \label{eq:CI19}
\end{equation}
for the anticommuting c-number field.
Then we have the differential equation
\begin{equation}
{{\delta \phi (x;\sigma )} \over {\delta \sigma (x')}}=
(-)^{\left| \phi \right| }\left[ {\phi (x;\sigma ),{\cal H}_{\rm int}[x']} \right]_c, \label{eq:CI20}
\end{equation}
which states that the interaction Hamiltonian density ${\cal H}_{\rm int} [x']$ is the
generating function to deform a space-like hypersurface for $\phi(x; \sigma)$. The equation (\ref{eq:CI20}) 
just corresponds to the  canonical equation (\ref{eq:IR1}). 
The integrability condition of (\ref{eq:CI20}) reduces to
\begin{equation}
\left[ {{\cal H}_{\rm int}[x'],{\cal H}_{\rm int}[x'']} \right]_c=0. \label{eq:CI21}
\end{equation}
However, the explicit verfication of the integrability condition
(\ref{eq:CI21}) is not necessary.
This is because,
unlike Koba's prescription, we already have the interaction Hamiltonian density ${\cal H}_{\rm int}[x']$
and the solution $\phi (x;\sigma)$ of the differential equation (\ref{eq:CI16}) or (\ref{eq:CI17})
in the integral form (\ref{eq:CI10}). 

As was shown in I, the four-dimensinal Poisson bracket is nothing but a representation of the symplectic
structure:
$$
\omega =-\int_\sigma  {d\Sigma _\nu }\bar \delta {\cal J}^\nu,
$$
where ${\cal J}^\nu$ is a symplectic current.
Then we can write (\ref{eq:CI20}) as
$$
{{\delta \phi (x;\sigma )} \over {\delta \sigma (x')}}=
(-)^{\left| \phi \right| }\omega \left( {{\bf X}_{\phi (x;\sigma )},{\bf X}_{{\cal H}_{\rm int}[x']}} \right).
$$
This means that the  canonical equation belongs to a representation in 
geometrical approach~\cite{CrWitt}. We may expect more progress in connecting our formalism with the symplectic
geometry~\cite{smplctc}. However, our interest is now in the correspondence (\ref{eq:CI20}) to the field equation
in quantum field theory rather than in the investigation of the symplectic geometry.
Thus we will proceed with our work in the form of the four-dimensional Poisson bracket. 

\section{The Yang--Feldman equation}

Let us go back to Gauss' theorem (\ref{eq:BD5}).
Put
$$
f^{\nu}[x'']=-{d(\partial )\Delta (x-x'') \!\cdot\! 
  \Bigl[ {\Gamma ^\nu (\partial '',\mathord{\buildrel{\lower3pt\hbox{$\scriptscriptstyle\leftarrow$}}\over {-\partial''} })
  \phi (x'')+ (-)^{\left| \phi \right| }\eta ^{-1}{S^\nu}^\dagger (x'')} \Bigr]}.
$$
We have immediately
\begin{eqnarray}
& & -\int_\Omega  {d^4x''\partial ''_\nu \left[ {d (\partial )\Delta (x-x'')
\cdot \Gamma ^\nu (\partial '',\mathord{\buildrel{\lower3pt\hbox{$\scriptscriptstyle\leftarrow$}}\over {-\partial''} } )
\phi (x'')+ (-)^{\left| \phi \right| }\eta ^{-1}{S^\nu }^\dagger (x'')} \right]}\nonumber \\
  & &=-\int_{\partial \Omega } {d\Sigma _\nu (x'')\left[ {d (\partial )\Delta (x-x'')\cdot
\Gamma ^\nu (\partial '',\mathord{\buildrel{\lower3pt\hbox{$\scriptscriptstyle\leftarrow$}}\over {-\partial''} } )
\phi (x'')+ (-)^{\left| \phi \right| }\eta ^{-1}{S^\nu }^\dagger (x'')} \right]}. \nonumber \\
& &\label{eq:YF1}
\end{eqnarray}
Using the canonical identity (\ref{eq:CI9}), the left-hand side ( LHS ) of (\ref{eq:YF1}) becomes
\begin{eqnarray}
{\rm LHS}&=&
-\int_\sigma ^{\sigma '} {d^4x''}\Bigl[  (-)^{\left| \phi \right| }\partial ''_\nu 
d (\partial )\Delta (x-x'')\cdot \eta ^{-1}{S^{\nu }}^\dagger (x'') \nonumber \\
& &\qquad\qquad- (-)^{\left| \phi \right| }d (\partial )\Delta (x-x'')\cdot 
\eta ^{-1}\left( {{{\partial ^R{\cal H}_{\rm int}[x'']} \over {\partial \phi (x'')}}} \right)^\dagger  \Bigr]. 
\nonumber \\
& &\label{eq:YF2}
\end{eqnarray}
Define
\begin{eqnarray}
\widetilde{ \phi} (x;\sigma )&=&
\phi (x'')-\int {d^4x''}\Bigl[ (-)^{\left| \phi \right| }\partial ''_\nu 
d (\partial )\Delta ^\sigma (x-x'')\cdot \eta ^{-1}{S^{\nu }}^\dagger (x'')\nonumber \\
& &{ }\qquad\qquad\qquad-(-)^{\left| \phi \right| }d (\partial )\Delta ^\sigma (x-x'')\cdot 
\eta ^{-1}\left( {{{\partial ^R{\cal H}_{\rm int}[x'']} \over {\partial \phi (x'')}}} \right)^\dagger \Bigr], \nonumber \\
& &\label{eq:YF3}
\end{eqnarray}
where $\Delta ^\sigma (x-x'')$ is the generalized Green's function:
\begin{eqnarray*}
\Delta ^\sigma (x-x'') &=&-\Delta^{\rm ret}(x-x'')\ {\rm if}\ x \ \hbox{is later than}\ \sigma \\
\Delta ^\sigma (x-x'') &=&-\Delta^{\rm adv}(x-x'')\ {\rm if}\ x \ \hbox{is earlier than}\ \sigma.
\end{eqnarray*}
The $\Delta^{\rm ret}(x-x'')$ and $\Delta^{\rm adv}(x-x'')$ 
represent the retarted and advanced Green's function, respectively.
Then (\ref{eq:YF2}) is simply written as 
\begin{equation}
{\rm LHS}=\widetilde{ \phi} (x;\sigma ')-\widetilde{ \phi} (x;\sigma ).
\label{eq:YF4}
\end{equation}
On the other hand, the right-hand side (RHS) of (\ref{eq:YF1}) gives
$$
{\rm RHS}=\phi (x;\sigma')-\phi (x;\sigma).
$$
Thus we get
\begin{eqnarray}
{{\delta \widetilde{ \phi} (x;\sigma )} \over {\delta \sigma (x')}}&=&
{{\delta \phi (x;\sigma )} \over {\delta \sigma (x')}}\nonumber\\
&=& -(-)^{\left| \phi \right| }\partial '_\nu d (\partial )\Delta (x-x')\cdot 
\eta ^{-1}{S^{\nu }}^\dagger (x') \nonumber \\
& &\qquad + (-)^{\left| \phi \right| }d (\partial )\Delta (x-x')\cdot 
\eta ^{-1}\left( {{{\partial ^R{\cal H}_{\rm int}[x']} \over {\partial \phi (x')}}} \right)^\dagger.
\label{eq:YF5}
\end{eqnarray}
Therefore the bubble differentiation of $\widetilde{\phi} (x;\sigma)$ equals to that of $\phi (x;\sigma)$.
We recall that equation (\ref{eq:CI10b}) is
given by the Klein-Gordon equation for an invariant delta function
and the half set of Hamilton's equations. Thus $\widetilde{ \phi} (x;\sigma)$ represents the integral form of the
field equation. Further, operating $\Lambda (\partial)$ from the left on (\ref{eq:YF3}), we get the
free-field equation
\begin{eqnarray}
\Lambda (\partial )\widetilde{ \phi} (x;\sigma )&=&
\Lambda (\partial )\phi (x)+ (-)^{\left| \phi \right| }\eta ^{-1}
\left[ {\partial _\nu {S^\nu }^\dagger (x)+
\left( {{{\partial ^R{\cal H}_{\rm int}[x]} \over {\partial \phi (x)}}} \right)^\dagger } \right]\nonumber \\
&=& (-)^{\left| \phi \right| }\eta ^{-1}
\left[ {\partial _n\Pi (x)+{{\partial ^R{\cal H}[x]} \over {\partial \phi (x)}}-
\partial _{t\mu }{{\partial ^R{\cal H}[x]} \over {\partial [\partial _{t\mu }\phi (x)]}}} \right]^\dagger \nonumber \\
&=&0 \label{eq:YF6},
\end{eqnarray}
the last step being verified when we make a foliation of spacetime into space-like sections
so that a piece of sections passing through the point $x$. 
We can really make the foliation of spacetime if we adopt 
succesive values of $c$ for a fixed function $g$ in (\ref{eq:BD1}).
Note that
\begin{eqnarray*}
& &\widetilde{ \phi} (x;\sigma )\to \phi ^{\rm in\ }(x) \qquad {\rm as} \ \sigma \to -\infty \\
& &\widetilde{ \phi} (x;\sigma )\to \phi ^{\rm out} (x) \qquad {\rm as} \ \sigma \to +\infty,
\end{eqnarray*}
where $\phi^{\rm in\ }(x)$ and $\phi^{\rm out} (x)$ 
represent an incomming and outgoing classical field, respectively.
This means that $\widetilde{ \phi} (x;\sigma )$ is the asymptotic field introduced by Yang and Feldman~\cite{YF}.
Thus we can identify equation (\ref{eq:YF3}) to be the general form of the Yang--Feldman equation.
Since both $\phi  (x;\sigma)$ and $\widetilde{ \phi} (x;\sigma )$ satisfy the free-field equation
and (\ref{eq:YF5}) holds, $\phi  (x;\sigma)$ is equivalent to the asymptotic field.

We wish finally to remark what could be manifest when a classical field is directly converted into a quantized field.
If we convert a classical field $\phi  (x')$ from the ordinary function to an operator $\widehat{\phi} (x')$ 
in the Heisenberg picture,
we have a quantized asymptotic field $\widehat{\phi}  (x;\sigma)$ of the form 
$\phi (x;\sigma)$ or $\widetilde{ \phi } (x;\sigma)$. 
The four-dimensional commutation or anticommutation relation 
of $\widehat{\phi} (x;\sigma)$ and $\widehat{\phi}(y;\sigma)$ is the usual free-field expression:
\begin{equation}
\left[ {\widehat{ \phi }  (x;\sigma ),\widehat{ \phi }(y;\sigma )} \right]_\pm =id (\partial _x)\Delta (x-y). \label{eq:YF8}
\end{equation}
Equation (\ref{eq:YF8}) holds for every space-like hypersurface $\sigma$, so there exsists a unitary operator
$U(\sigma,\sigma')$~\cite{YF} such that
\begin{equation}
\widehat{ \phi } (x;\sigma ')=U^{-1}(\sigma ,\sigma ')\widehat{ \phi } (x;\sigma )U(\sigma ,\sigma '). \label{eq:YF9}
\end{equation}
It follows that
\begin{equation}
i{{\delta \widehat{ \phi }  (x;\sigma )} \over {\delta \sigma (x')}}
=\left[ {\widehat{ \phi }  (x;\sigma ),\widehat{ H}(x'/ \sigma )} \right]_\pm \label{eq:YF10} 
\end{equation}
where
$$
\widehat{ H}(x'/ \sigma )\equiv 
i\left( {{{\delta U(\sigma ,\sigma ')} \over {\delta \sigma (x')}}} \right)_{\sigma =\sigma '},
$$ 
and $x'/ \sigma$ is the abbreviation for $x'$ on $\sigma$.
Note that the $S$-matrix ${\bf S}$ in the Heisenberg picture is defined by the unitary operator $U(\sigma,\sigma')$:
$$
{\bf S}=U(\infty ,-\infty )=\cdots U(\sigma _0,\sigma _{-1})U(\sigma _1,\sigma _0)U(\sigma _2,\sigma _1) \cdots,
$$
where $\cdots \sigma_1, \ \sigma_0, \ \sigma_{-1} \cdots$ denote 
an infinite sequence of space-like hypersurfaces which proceed steady into past.
Yang and Feldman investigated several examples of interactions and found
$\widehat{ H}(x'/ \sigma )$ for given  ${\widehat{ \phi} (x;\sigma )}$. We can verify each of
$\widehat{ H}(x'/ \sigma )$ calculated by Yang and Feldman has the same form of 
${\cal H}_{\rm int}[x']$ given by $T^{\mu\nu}[x']$.
On the other hand, if we classicize
their examples of $\widehat{H}(x'/ \sigma)$ to the function $H(x'/ \sigma)$
and substitute $H(x'/ \sigma)$
for ${\cal H}_{\rm int}[x']$, the $H(x'/\sigma)$ satisfy (\ref{eq:CI18}) or (\ref{eq:CI19}).
It follows that
\begin{equation}
{{\delta \phi (x;\sigma )} \over {\delta \sigma (x')}}
=(-)^{\left| \phi \right| }\left[ {\phi (x;\sigma ), H(x'/ \sigma )} \right]_c. \label{eq:YF11}
\end{equation}
Thus the canonical equation (\ref{eq:YF11}) corresponds to the differential form of 
the Yang-Feldman equation (\ref{eq:YF10}) as for their examples.
It is not evident that, in general, any $H(x'/ \sigma)$ satisfies (\ref{eq:CI18}) or
(\ref{eq:CI19}), and  exactly coincides with ${\cal H}_{\rm int}[x']$.
The question what kind of interactions necessarily satisfies (\ref{eq:CI18}) or (\ref{eq:CI19})
requires further investigation.

The application of our formulation is presented for quantum chromodynamics (QCD) in a separate paper~\cite{oz3}.
\newpage
\section*{Appendix}

We can express $\phi  (x;\sigma)$ in terms of canonical variables on $\sigma$
\begin{eqnarray*}
\phi (x;\sigma )&=&
\int_\sigma  d\Sigma [-d (\partial )\Delta (x-x'')\cdot  (-)^{\left| \phi \right| }\eta ^{-1}\Pi^\dagger (x'') \\
& &\qquad\quad +n_\nu (x'')\Lambda ^{\nu \lambda }\partial ''_\lambda d(\partial )\Delta (x-x'')\cdot \phi (x'')].
\end{eqnarray*}
Then the four-dimensional Poisson bracket of $\phi  (x;\sigma)$ and ${\cal H}_{\rm int}[x']$
reads
\begin{eqnarray*}
\left[ {\phi (x;\sigma ),{\cal H}_{\rm int}[x']} \right]_c
&=&[n_\nu (x')\Lambda ^{\nu \lambda }\partial '_\lambda d(\partial )\Delta (x-x') \\
& &{ }-d(\partial )\Delta (x-x')\!\cdot\!  (-)^{\left| \phi \right|}
  \!\left( {{{\partial ^R} \over {\partial \phi (x')}}-\partial' _{t\mu }
  {{\partial ^R} \over {\partial [\partial' _{t\mu }\phi (x')]}}} \right)\eta ^{-1}\Pi ^\dagger (x')] \\
& &\quad\times \left( {{{\partial ^R{\cal H}_{\rm int}[x']} \over {\partial \Pi (x')}}
  +\mathord{\buildrel{\lower3pt\hbox{$\scriptscriptstyle\leftarrow$}}\over {\partial ^{\prime}}} _{t\mu}
  {{\partial ^R{\cal H}_{\rm int}[x']} \over {\partial [\partial '_{t\mu }\Pi (x')]}}} \right) \\
& &{ }+[ d (\partial )\Delta (x-x')\cdot \left( {{{\partial ^R} \over {\partial \Pi (x')}}
  -\partial' _{t\mu }{{\partial ^R} \over {\partial [\partial' _{t\mu }\Pi (x')]}}} \right)\eta^{-1}\Pi^\dagger (x')] \\
& &\quad\times \left( {{{\partial ^R{\cal H}_{\rm int}[x']} \over {\partial \phi (x')}}
  +\mathord{\buildrel{\lower3pt\hbox{$\scriptscriptstyle\leftarrow$}}\over {\partial^{\prime}}} _{t\mu}
  {{\partial ^R{\cal H}_{\rm int}[x']} \over {\partial [\partial '_{t\mu }\phi (x')]}}} \right). \qquad\qquad\quad (A.1)
\end{eqnarray*}

The half set of Hamilton's equations  
$$
(-)^{\left| \phi \right| +1}\partial ''_n\phi (x'')-\partial ''_{t\mu }
{{\partial ^R{\cal H}[x'']} \over {\partial [\partial'' _{t\mu }\Pi (x'')]}}
+{{\partial ^R{\cal H}[x'']} \over {\partial \Pi (x'')}}=0 \eqno(A.2)
$$
must be treated carefully. For the c-number field
satisfying a second-order field equation, equation $(A.2)$ 
does not give the field equation but leads to a trivial identity: 0=0. For the anticommuting
c-number field satisfying a first-order field equation, 
equation $(A.2)$ correctly gives the field equation. The kinetic terms of the 
field equation for the anticommuting c-number field becomes
$$
\Lambda (\partial '')\phi (x'')
=\zeta \left[ {(-)^{\left| \phi \right| +1}\partial ''_n\phi (x'')-\partial ''_{t\mu }
{{\partial ^R{\cal H}_0[x'']} \over {\partial [\partial'' _{t\mu }\Pi (x'')]}}
+{{\partial ^R{\cal H}_0[x'']} \over {\partial \Pi (x'')}}} \right].
$$
Then we have a canonical identity
\begin{eqnarray*}
0&=&d (\partial )\Lambda (\partial )\Delta (x-x'')\cdot \phi (x'') \\
& &{ }-d (\partial )\Delta (x-x'')\cdot 
\zeta \left[ {(-)^{\left| \phi \right| +1}\partial ''_n\phi (x'')-\partial ''_{t\mu }
{{\partial ^R{\cal H}[x'']} \over {\partial [\partial''_{t\mu }\Pi (x'')]}}
+{{\partial ^R{\cal H}[x'']} \over {\partial \Pi (x'')}}} \right] \\
&=&
-\partial ''_\nu \left[ {d (\partial )\Delta (x-x'')
\Gamma ^\nu (\partial '',-\mathord{\buildrel{\lower3pt\hbox{$\scriptscriptstyle\leftarrow$}}\over {\partial''} })
\phi (x'')} \right] \\
& &{ }
-d (\partial )\Delta (x-x'')\cdot 
\zeta \left[ {{{\partial ^R{\cal H}_{\rm int}[x'']} \over {\partial \Pi (x'')}}-\partial ''_{t\mu }
{{\partial ^R{\cal H}_{\rm int}[x'']} \over {\partial [\partial'' _{t\mu }\Pi (x'')]}}} \right]. \qquad\qquad\quad(A.3)
\end{eqnarray*}
Substituting (A.1)--(A.3) we have (\ref{eq:CI17}).

\end{document}